\documentclass[a4paper]{article}
\usepackage{amsmath,amsthm,amssymb}

\usepackage[table]{xcolor}

\usepackage{xcolor}

\usepackage{wrapfig}

\usepackage{multirow}
\usepackage{rotating}
\usepackage{verbatim} 
\usepackage{booktabs}
\usepackage{graphicx}
\usepackage{caption}
\usepackage{subcaption}
\usepackage{shellesc}

\usepackage{parskip}
\usepackage{listings}
\usepackage{color}

\usepackage[frozencache,cachedir=.]{minted}

\usepackage{url}

\usepackage{algorithmic,algorithm}

\usepackage{authblk}

\providecommand{\keywords}[1]
{
  \small	
  \textbf{Keywords---} #1
}

\title{Migrating CUDA to oneAPI: A Smith-Waterman Case Study}
\author[1]{Manuel Costanzo}
\author[1]{Enzo Rucci}
\author[2]{Carlos Garcia Sanchez}
\author[1]{Marcelo Naiouf}
\author[2]{Manuel Prieto-Matias}

\affil[1]{III-LIDI, Facultad de Inform\'atica,
Universidad Nacional de La Plata - CIC\\
La Plata, Buenos Aires, Argentina\\
 \authorcr
\{mcostanzo,erucci,mnaiouf\}@lidi.info.unlp.edu.ar} 

\affil[2]{Dpto. Arquitectura de Computadores y Automática, Universidad Complutense de Madrid.  Madrid (28040), España\\
\authorcr
\{garsanca,mpmatias\}@dacya.ucm.es}

\date{{March 1, 2022}}

\begin{document}

\maketitle              

\begin{abstract}
To face the programming challenges related to heterogeneous computing, Intel recently introduced oneAPI, a new programming environment that allows code developed in Data Parallel C++ (DPC++) language to be run on different devices such as CPUs, GPUs, FPGAs, among others. 
To tackle CUDA-based legacy codes, oneAPI provides a compatibility tool (\texttt{dpct}) that  facilitates  the  migration  to  DPC++. Due to the large amount of existing CUDA-based software in the bioinformatics context, this 
 paper presents our experiences porting \textit{SW\#db}, a well-known sequence alignment tool, to DPC++ using \texttt{dpct}. From the experimental work, it was possible to prove the  usefulness of \texttt{dpct} for SW\#db code migration and the cross-GPU vendor, cross-architecture portability of the migrated DPC++ code. In addition, the performance results showed that the migrated DPC++ code reports similar efficiency rates to its CUDA-native counterpart or even better in some tests (approximately +5\%).

\end{abstract}

\begin{center}
\texttt{This version of the contribution has been accepted for publication, after peer review (when applicable) but is not the Version of Record and does not reflect post-acceptance improvements, or any corrections. The
Version of Record is available online at: \url{https://doi.org/10.1007/978-3-031-07802-6\_9}. Use of this
Accepted Version is subject to the publisher’s Accepted Manuscript terms of use
\url{https://www.springernature.com/gp/open-research/policies/accepted-manuscript-terms”}}
\end{center}

\clearpage

\section{Introduction}
\label{sec:intro}

At present, heterogeneous computing and massively parallel architectures have proven to be an effective strategy for maximizing the performance and energy efficiency of computing systems~\cite{zahran2017heterogeneous}
. That is the main reason why the programmers typically rely on a variety of hardware, like CPU, GPU, FPGAs, and other kinds of accelerators.  This raises the need for specialized libraries, tools, and APIs that increase the programming cost and complexity, and complicate future code maintenance and extension. 

On the one hand, Khronos Group has proposed SYCL~\footnote{\url{https://www.khronos.org/registry/SYCL/specs/sycl-2020/pdf/sycl-2020.pdf}}, an open standard, to face some of the programming issues related to
heterogeneous computing. Although SYCL shares some characteristics with OpenCL (such as being royalty-free and cross-platform), it can actually be seen as an improved, high-level version of the latter. SYCL is an abstraction layer that enables code for heterogeneous systems to be written using standard, single-source C++ host code
including accelerated code expressed as functions or \textit{kernels}. SYCL implementations are often based on OpenCL, but also have the flexibility to use other backends like CUDA or OpenMP. Furthermore, SYCL features asynchronous task graphs, 
buffers defining location-independent storage, 
interoperability with OpenCL, among other characteristics oriented to increase productivity~\cite{SYCL-spec,EarlyExperimentsUsingSYCL-FPGA}. 

On the other hand, Intel recently introduced the \textit{oneAPI} programming ecosystem that provides a unified programming model for a wide range of hardware architectures. The core of the oneAPI environment is a simplified language to express parallelism in heterogeneous platforms, named Data Parallel C++ (DPC++), which can be summarized as C++ with SYCL. In addition, oneAPI also comprises a runtime, a set of domain-focused libraries and supporting tools~\cite{PortingLegacyCUDAtoOneAPI}.

In this scenario, GPUs can be considered the dominant accelerator and CUDA is the most popular programming language for them nowadays~\cite{CUDApopular}. Bioinformatics and Computational Biology are some of the communities that have been exploiting GPUs for more than two decades~\cite{GPUsInBioinformatics2016}. Lots of GPU implementations can be found in sequence alignment~\cite{SandesReview2016}, molecular dynamics~\cite{loukatou2014molecular}, molecular docking~\cite{ohue2014megadock}, prediction and searching of molecular structures~\cite{mrozek2014parallel}, among other application areas. 
Even though some applications achieve a better performance with CUDA, their portability to other architectures is strongly restricted due to their proprietary nature.

To tackle CUDA-based legacy codes, oneAPI provides a compatibility tool (\texttt{dpct}) that facilitates the migration to the SYCL-based DPC++ programming language. A few preliminary studies assessing \texttt{dpct} usefulness can be found in simulation~\cite{PortingLegacyCUDAtoOneAPI}, math~\cite{tsai2021porting,costanzo2021early}, and cryptography~\cite{xjoin_oneapi}; however, to the best of our knowledge, no study assess their utility in Bioinformatics arena. In this paper, we present our experiences porting a biological software tool to DPC++ using \texttt{dpct}. In particular, we have selected \textit{SW\#db}~\cite{swsharpdb}: a CUDA-based, memory-efficient implementation of the Smith-Waterman  (SW) algorithm, that can be used either as a stand-alone application or a library.
Our contributions are:

\begin{itemize}
\item An analysis of the \texttt{dpct} effectiveness for the CUDA-based SW\#db migration, including a detailed summary of the porting steps that required manual modifications.

\item An analysis of the DPC++ code's portability, considering different target platforms and vendors (Intel CPUs and GPUs; NVIDIA GPUs).

\item A comparison of the performance on different hardware architectures (Intel CPUs and GPUs; NVIDIA GPUs).

\end{itemize}

This work can be considered the starting point for a more exhaustive evaluation exploration of CUDA-based biological tool migration to oneAPI. The remaining sections of this article are organized as follows: in Section~\ref{sec:back}, the background is presented. Next, in Section~\ref{sec:imple}, the migration process is described, and in Section~\ref{sec:results}, the experimental work carried out is detailed and the results obtained are analyzed. Finally, in Section~\ref{sec:conc}, the conclusions and possible lines of future work are presented.







\section{Background}
\label{sec:back}

\subsection{The oneAPI Programming Ecosystem}

oneAPI~\footnote{\url{https://www.oneapi.com/}} is a unified programming model for application development that can be used on different architectures, such as CPUs, GPUs, and even FPGAs. oneAPI seeks to facilitate the hard task of developing applications on a different set of hardware. 
Using oneAPI, the coding task can be performed at various levels: (1) invoking one of the multiple optimized libraries (oneMKL, oneDAL, oneVPL, etc) that takes advantage of offloading technology in a transparent way to the programmer, or (2) direct programming using the SYCL heterogeneous programming language supported by the Data Parallel C++ (DPC++) language. The DPC++ programming language (supported by Intel's \texttt{dpcpp} compiler) combines the C++ language with SYCL, allowing the same source code to be compiled and executed across different accelerators. 

oneAPI comprises several programming tools and one of the most interesting considering code migration is a compatibility one named as \texttt{dpct}. This tool converts applications written in the proprietary CUDA language to SYCL. According to Intel, this tool  automatically migrates 80\%-90\% of the original CUDA code to SYCL. In addition, regarding non-ported code, \texttt{dpct} inlines comments (through warning messages) that help the programmer to migrate and tune the final DPC++ code.~\cite{ParallelUniverseOneAPI}.

The migration process consists of 3 stages:
\begin{enumerate}
    \item Run the \texttt{dpct} tool that performs the automatic code migration.
    \item Modify the migrated code attending all \texttt{dpct} warnings to reach a first, executable version following the diagnostics reference\footnote{Diagnostics Reference of Intel® DPC++ Compatibility Tool available at: \url{https://software.intel.com/content/www/us/en/develop/documentation/intel-dpcpp-compatibility-tool-user-guide/top/diagnostics-reference.html}}.
    \item Verify the correctness and efficiency of the resulting oneAPI program and make the necessary modifications accordingly.
\end{enumerate}

\subsection{Smith-Waterman Algorithm}

\label{sec:SW-Algorithm}

This algorithm was proposed by Smith and Waterman~\cite{Smith1981} to obtain the optimal local alignment between two biological sequences. SW employs a dynamic programming approach and presents quadratic time
and space complexities. Furthermore, it has been used as
the basis for many subsequent algorithms and is often employed as a benchmark when comparing different alignment techniques~\cite{Hasan2011}.

The SW algorithm can be used to compute (a) pairwise alignments (one-to-one) or (b) database similarity searches (one-to-many). Both cases
have been parallelized in the literature. In case (a),
a single SW matrix is calculated and all Processing Elements (PEs) work collaboratively (\textit{intra-task parallelism}). Due to inherent data dependencies, neighbour PEs communicate in order to exchange border elements.  In case (b), multiple SW matrices are calculated simultaneously without
communication between the PEs (\textit{inter-task parallelism})~\cite{SandesReview2016}.

\subsection{SW\#}

SW\# is a tool to compute biological sequence alignments that can be used as an API-based library or as a standalone  command-line executable~\cite{swsharp}.
It is considered a versatile tool since it works with both protein and DNA sequences, being able to compute pairwise alignments as well as database similarity searches. 

SW\#db is the package  for fast exact similarity searches, 
which works by simultaneously utilizing CPU and GPU(s). The GPU part is based on CUDA and follows both inter-task and intra-task parallelism approaches (depending on the sequence lenght). On its behalf, CPU just exploits inter-task paralleism through multithreading and SIMD instructions~\footnote{In particular, it makes use of the OPAL library for the CPU part~\url{https://github.com/Martinsos/opal}}.
Through dynamic work distribution and dynamic communication between the CPU-GPU, SW\#db significantly reduces execution time.




\section{Implementation}
\label{sec:imple}

\subsection{Differences between CUDA and DPC++}

Before migrating a code from CUDA to oneAPI, some differences should be considered.

\subsubsection{Memory model:}

on the one hand, CUDA provides two different types of memory model:
\begin{enumerate}
    \item Conventional model: it is mandatory to specify all memory operations between the CPU and GPU.
    \item Unified Memory (UM): introduced in CUDA 6, this model creates a shared memory pool between the CPU and GPU, where both access the data through pointers in a transparent manner to the programmer.
\end{enumerate}

On the other hand, oneAPI offers three abstractions for managing memory:
\begin{enumerate}
    \item Buffers: they are data abstractions that represent one or more objects of a given C++ language type. Buffers represent data objects rather than specific memory addresses, so the same buffer can be allocated to several different memory locations on different devices, or even on the same device, for performance reasons.
    \item Images: they are a special type of buffer created especially for image processing. They include support for special image formats, image reading through sampling objects, among others.
    \item Unified Shared Memory (USM): it consists of creating a unified virtual memory space where pointers are shared between the CPU and the device (similar to the CUDA UM).    
\end{enumerate}

\subsubsection{Verbosity:}

in DPC++, all variables used within a kernel must be declared and explicitly sent to the functions, as well as other aspects that in CUDA are not mandatory. On the contrary, in CUDA it is possible to indicate the variables that you want to send to the device and  implicitly use them in the kernels. These issues may cause the oneAPI code to be longer than its CUDA counterpart.

\subsection{Migrating CUDA Codes to DPC++}

As a general, \texttt{dpct} is not able to generate a fully functional DPC++ code. Thus, it is required to perform hand-tunned adaptations. However, the \texttt{dpct} tool reports list of warnings which facilitates successful refactoring. 

\subsubsection{Warnings generated by \texttt{dpct}:} these warnings vary among simple recommendations (i.e to improve the performance) to more complex issues, such as fragments code not successfully migrated.

This section will detail the messages reported by the migration \texttt{dpct} tool when porting the SW\# and the manual adaptation made to obtain the final DPC++ code. 

\begin{lstlisting}[
  breaklines,
  frame=single,
]
DPCT1003: Migrated API does not return error code. (*, 0) is inserted. You may need to rewrite this code

DPCT1009: SYCL uses exceptions to report errors and does not use the error codes. The original code was commented out and a warning string was inserted. You need to rewrite this code.
\end{lstlisting}

Both warnings occur when using native CUDA functions, such as CUDA error codes (Fig.~\ref{fig:cuda-cuda_safe_call}). Since \texttt{dpct} cannot translate them, it modifies the code to still keep it functional (Fig.~\ref{fig:oneapi-cuda_safe_call}). Generally, this technique is used when exchanging data with the device. Figs.~\ref{fig:cuda-cuda_safe_call}) and~\ref{fig:oneapi-cuda_safe_call}) show memory allocations on the GPU using CUDA and oneAPI, respectively. By default, \texttt{dpct} tries to use the \texttt{USM} model because it produces less volume of code and allows \texttt{dpct} to support more memory-related APIs.

\begin{figure}
\centering
\begin{subfigure}{.49\textwidth}
  \centering
\begin{minted}{c}
size_t valuesSize = 
    databaseLen * sizeof(double);
double* valuesGpu;

CUDA_SAFE_CALL(
    cudaMalloc(
        &valuesGpu, valuesSize
    )
);
\end{minted}
  \caption{CUDA}
  \label{fig:cuda-cuda_safe_call}
\end{subfigure}%
\begin{subfigure}{.49\textwidth}
  \centering
\begin{minted}{c}
size_t valuesSize = 
    databaseLen * sizeof(double);
double* valuesGpu;
    
CUDA_SAFE_CALL((
 valuesGpu = (double *)sycl::malloc_device(
    valuesSize, dpct::get_default_queue()),
0));
\end{minted}
  \caption{oneAPI}
  \label{fig:oneapi-cuda_safe_call}
\end{subfigure}
\caption{\texttt{CUDA\_SAFE\_CALL} example}
\label{fig:test}
\end{figure}

\begin{lstlisting}[
  breaklines,
  frame=single,
]
DPCT1005: The SYCL device version is different from CUDA Compute Compatibility. You may need to rewrite this code.
\end{lstlisting}

This problem is related to the previous one and appears when querying for intrinsic CUDA attributes. While \texttt{dpct} can obtain information from the GPU, such as the number of registers, maximum memory size, among others~\footnote{\url{https://docs.oneapi.io/versions/latest/dpcpp/iface/device.html}}, some CUDA-proprietary attributes (e.g. CUDA driver information) are not translatable. Fig.~\ref{fig:cuda-properties} shows that, in the original code, the number of CUDA blocks and threads depends on the driver version. Fig.~\ref{fig:oneapi-properties} presents the migrated code, showing that it is possible to obtain information about GPU properties, with the exception of those specific to CUDA.

\begin{figure}
\centering
\begin{subfigure}{.49\textwidth}
  \centering
\begin{minted}{c}
cudaDeviceProp properties;
cudaGetDeviceProperties(
    &properties, card
);

int threads;
int blocks;
if (properties.major < 2) {
    threads = THREADS_SM1;
    blocks = BLOCKS_SM1;
} else {
    threads = THREADS_SM2;
    blocks = BLOCKS_SM2;
}
\end{minted}
  \caption{CUDA}
  \label{fig:cuda-properties}
\end{subfigure}%
\begin{subfigure}{.49\textwidth}
  \centering
\begin{minted}{c}
dpct::device_info properties;

dpct::dev_mgr::instance()
    .get_device(card)
    .get_device_info(properties);
    
int threads;
int blocks;

if (false) {
    threads = THREADS_SM1;
    blocks = BLOCKS_SM1;
} else {
    threads = THREADS_SM2;
    blocks = BLOCKS_SM2;
}
\end{minted}
  \caption{oneAPI}
  \label{fig:oneapi-properties}
\end{subfigure}
\caption{Querying device properties}
\label{fig:properties}
\end{figure}

\begin{lstlisting}[
  breaklines,
  frame=single,
]
DPCT1049: The workgroup size passed to the SYCL kernel may exceed the limit. To get the device limit, query info::device::max_work_group_size. Adjust the workgroup size if needed
\end{lstlisting}

To run the CUDA kernel, both block and thread sizes must be configured; however, each device have a different size limit. \texttt{dpct} alerts the programmer that the migrated code may exceed the maximum work-group limit that the underlying architecture supports. In addition, it recommends adjusting the code if necessary. Fig.~\ref{fig:cuda-workgroups} shows how to run the kernel in CUDA, while Fig.~\ref{fig:oneapi-workgroups} shows the DPC++ counterpart.

\begin{figure}
\centering
\begin{subfigure}{.49\textwidth}
  \centering
\begin{minted}{c}
solveShort<<<blocks, threads>>>(...);
\end{minted}
  \caption{CUDA}
  \label{fig:cuda-workgroups}
\end{subfigure}%
\begin{subfigure}{.49\textwidth}
  \centering
\begin{minted}{c}
dpct::get_default_queue()
    .submit([&](sycl::handler &cgh) {
 ...
cgh.parallel_for(
  sycl::nd_range<3>
    (sycl::range<3>(1, 1, blocks) *
    sycl::range<3>(1, 1, threads),
    sycl::range<3>(1, 1, threads)),
  [=](sycl::nd_item<3> item_ct1) {
    solveShort(...);
  });
});
\end{minted}
  \caption{oneAPI}
  \label{fig:oneapi-workgroups}
\end{subfigure}
\caption{Kernel launch with dynamic work-group size}
\label{fig:workgroups}
\end{figure}

\begin{lstlisting}[
  breaklines,
  frame=single,
]
DPCT1065: Consider replacing sycl::nd_item::barrier() with sycl::nd_item::barrier(sycl::access::fence_space::local_space) for better performance if there is no access to global memory
\end{lstlisting}

On this situation, \texttt{dpct} recommends the programmer to use an additional parameter when synchronizing threads within the kernel as long as no global memory is used. By default, the tool does not automatically optimize this issue because it cannot discern whether this memory is being used. An example of the CUDA thread synchronization and the migrated oneAPI code can be seen in the Figs.~\ref{fig:cuda-sync} and~\ref{fig:oneapi-sync}, respectively.

\begin{figure}
\centering
\begin{subfigure}{.49\textwidth}
  \centering
\begin{minted}{c}
...
__syncthreads();
...
\end{minted}
  \caption{CUDA}
  \label{fig:cuda-sync}
\end{subfigure}%
\begin{subfigure}{.49\textwidth}
  \centering
\begin{minted}{c}
...
item_ct1.barrier();

//DPCT recommends adding the following parameter.
/*item_ct1.barrier(
    sycl::access::fence_space::local_space
);*/
...
\end{minted}
  \caption{oneAPI}
  \label{fig:oneapi-sync}
\end{subfigure}
\caption{Thread synchronization}
\label{fig:sync}
\end{figure}

\begin{lstlisting}[
  breaklines,
  frame=single,
]
DPCT1084: The function call has multiple migration results in different template instantiations that could not be unified. You may need to adjust the code.
\end{lstlisting}

In CUDA, generic functions are a common way to reduce code size, since they permit code reusing for data of different type. Although oneAPI supports this programming feature, it cannot automatically port this kind of code due to the multiplicity of possible migration options. Fig.~\ref{fig:cuda-generica} shows a CUDA example where instructions depend on the type of parameter sent to the kernel function. Fig.~\ref{fig:oneapi-generica} shows the corresponding migrated code.

\begin{figure}[t]
\centering
\begin{subfigure}{.49\textwidth}
  \centering
\begin{minted}{c}
class SubScalarRev {
public:
  __device__ int operator()(...) {
    ...
  }
};

class SubVector {
public:
  __device__ int operator()(...) {
    ...
  }
};

template <class Sub>
__global__ static void solveLong(
..., Sub sub) {
    sub(...);
}


solveLong<<<blocks, threads>>>(
    ..., SubScalarRev()
);

solveLong<<<blocks, threads>>>(
 ..., SubVector());
\end{minted}
  \caption{CUDA}
  \label{fig:cuda-generica}
\end{subfigure}%
\begin{subfigure}{.49\textwidth}
  \centering
\begin{minted}{c}
class SubScalarRev {
public:
  int operator()(...) {
   ...
  }
};

class SubVector {
public:
  int operator()(...) {
      ...
  }
};

template <class Sub>
static void solveLong(..., Sub sub) {
    sub(...);                         
}

cgh.parallel_for(
  sycl::nd_range<3>
    (sycl::range<3>(1, 1, blocks) *
    sycl::range<3>(1, 1, threads),
    sycl::range<3>(1, 1, threads)),
  [=](sycl::nd_item<3> item_ct1) {
    solveLong(..., SubScalarRev());
  });
});

cgh.parallel_for(
  sycl::nd_range<3>
    (sycl::range<3>(1, 1, blocks) *
    sycl::range<3>(1, 1, threads),
    sycl::range<3>(1, 1, threads)),
  [=](sycl::nd_item<3> item_ct1) {
    solveLong(..., SubVector());
  });
});
\end{minted}
  \caption{oneAPI}
  \label{fig:oneapi-generica}
\end{subfigure}
\caption{Generic functions}
\label{fig:generica}
\end{figure}

\begin{lstlisting}[
  breaklines,
  frame=single,
]
DPCT1059: SYCL only supports 4-channel image format. Adjust the code.
\end{lstlisting}

In CUDA, texture memory variables can contain from 1 to 4 channels. In SYCL, texture memory is accessed through images. As it was reported by the \texttt{dpct} warning, SYCL only supports the use of 4-channel images, so the programmer must adapt the parts of the code where images of different sizes are used. In Fig.~\ref{fig:cuda-textura}  a 1-channel texture variable is declared in CUDA (placed in the device) and finally a data is read from it. Fig.~\ref{fig:oneapi-textura} presents a possible adjustment to corresponding code to convert a texture 1-channel variable to an equivalent 4-channel one. As it can be seen, this conversion requires to modify the indexes through which the memory is accessed to obtain the correct data. In that sense, a 2-bit right shift  (equivalent to DIV 4) combined with a logical \texttt{AND 3} operation (equivalent to MOD 4) must be performed in the corresponding read operation.

\begin{figure}
\centering
\begin{subfigure}{.49\textwidth}
  \centering
\begin{minted}{c}
texture<char> colTexture;

int colSize = colsGpu * sizeof(char);
char *colGpu;
cudaMalloc(&colGpu, colSize);
cudaMemcpy(colGpu, colCpu, 
    colSize, TO_GPU);
cudaBindTexture(NULL, colTexture, 
    colGpu, colSize);

char v = tex1Dfetch(colTexture, 10);
\end{minted}
  \caption{CUDA}
  \label{fig:cuda-textura}
\end{subfigure}%
\begin{subfigure}{.49\textwidth}
  \centering
\begin{minted}{c}
// Migración DPCT
//dpct::image_wrapper<char, 1> colTexture;

dpct::image_wrapper<sycl::char4, 1> colTexture;

int colSize = colsGpu * sizeof(char);
char* colGpu;

colGpu = (char *)sycl::malloc_device(
    colSize, dpct::get_default_queue());

dpct::get_default_queue()
    .memcpy(colGpu, colCpu, colSize).wait();

colTexture.attach(colGpu, colSize);

// DIV 4 y MOD 3
char v = colTexture.read(10 >> 2)[10 & 3];
\end{minted}
  \caption{oneAPI}
  \label{fig:oneapi-textura}
\end{subfigure}
\caption{4-channel texture memory}
\label{fig:textura}
\end{figure}

\subsubsection{Runtime and results check:}

once the code compiles correctly, it must be verified that there are no execution errors and that the results obtained are correct. In this case, although the oneAPI program compiled correctly, the following runtime error appeared:

\begin{lstlisting}[
  breaklines,
  frame=single,
]
For a 1D/2D image/image array, the width must be a Value >= 1 and <= CL_DEVICE_IMAGE2D_MAX_WIDTH
\end{lstlisting}

This error appears because SYCL images have a limited size, being the maximum size of the 1D images (vectors) smaller than their 2D counterparts (matrices). To solve this issue, this image object must be converted to another DPC++ memory abstraction: buffers or USM. We have chosen USM because the required modifications were simpler compared to the other option.

Fig.~\ref{fig:cuda-matrix_usm} shows how a 2-level texture memory is allocated on the GPU, while Fig.~\ref{fig:oneapi-matrix_usm} illustrates how to use USM to send a the array to the device. In that sense, the read mechanism also changes, both in CUDA (Fig.~\ref{fig:cuda-matrix_read}) and in DPC++.

\begin{figure}
\centering
\begin{subfigure}{.49\textwidth}
  \centering
\begin{minted}{c}
texture<int, 2, 
  cudaReadModeElementType> seqsTexture;

cudaArray *sequencesGpu;
cudaChannelFormatDesc channel = 
  seqsTexture.channelDesc;
cudaMallocArray(&sequencesGpu, 
    &channel, sequencesCols, sequencesRows);
cudaMemcpyToArray(sequencesGpu, 0, 0, 
    sequences, sequencesSize, TO_GPU);
cudaBindTextureToArray(
    seqsTexture, sequencesGpu);
\end{minted}
  \caption{CUDA}
  \label{fig:cuda-matrix_usm}
\end{subfigure}%
\begin{subfigure}{.49\textwidth}
  \centering
\begin{minted}{c}
static int *seqsGpu;

seqsGpu = (int *)sycl::malloc_device(
 sequencesCols * sequencesRows * sizeof(int), 
 dpct::get_default_queue());

dpct::get_default_queue()
 .memcpy(seqsGpu, sequences, sequencesCols * 
        sequencesRows * sizeof(int))
  .wait();
\end{minted}
  \caption{oneAPI}
  \label{fig:oneapi-matrix_usm}
\end{subfigure}
\caption{CUDA 2-D texture memory adaptation using DPC++ USM}
\label{fig:matrix_usm}
\end{figure}

\begin{figure}
\centering
\begin{subfigure}{.49\textwidth}
  \centering
\begin{minted}{c}
int columnCodes = tex2D(
  seqsTexture, colOff, j + rowOff);
\end{minted}
  \caption{CUDA}
  \label{fig:cuda-matrix_read}
\end{subfigure}%
\begin{subfigure}{.49\textwidth}
  \centering
\begin{minted}{c}
int columnCodes = 
    seqsGpu[(j + rowOff) * sequencesCols + colOff];
\end{minted}
  \caption{oneAPI}
  \label{fig:oneapi-matrix_read}
\end{subfigure}
\caption{Data accessing in 2D array}
\label{fig:matrix_read}
\end{figure}

After the DPC++ program executed successfully, different tests were performed and their results were verified to ensure that they were equivalent to those of the original CUDA code.
\section{Experimental Results}
\label{sec:results}

\subsection{Experimental design}

All tests were carried out using the platforms described in Table~\ref{tab:platforms}. oneAPI and CUDA versions are 2022.0 and 11.5, respectively, and to run DPC++ codes on NVIDIA GPUs, we build a DPC++ toolchain with support for NVIDIA CUDA, as it is not supported by default on oneAPI~\footnote{https://intel.github.io/llvm-docs/GetStartedGuide.html}. The performance was evaluated by carrying out similar experiments to those in previous works~\cite{OSWALD16,rucci2019swimm2}, searching 20 query protein sequences against the well-known UniProtKB/Swiss-Prot database (release 2021\_04)\footnote{Swiss-Prot: ~\url{https://www.uniprot.org/downloads}}:

\begin{itemize}
    \item The input queries range in length from 144 to 5478, and they were extracted from the Swiss-Prot database (accession numbers: P02232, P05013, P14942, P07327, P01008, P03435, P42357, P21177, Q38941, P27895, P07756, P04775, P19096, P28167, P0C6B8, P20930, P08519, Q7TMA5, P33450, and Q9UKN1). 
    \item The database contains 204173280 amino acid residues in 565928 sequences with a maximum length of 35213. 
    \item BLOSUM62 and 10(2) were set as the scoring matrix and gap insertion (extension) penalty, respectively.
\end{itemize}

As SW\#db is a hybrid CPU-GPU software, just a single thread was configured at CPU level (flag \texttt{T=1}) to minimize its impact on the overall performance. Besides, different work-group~\footnote{A DPC++ work-group is equivalent to a CUDA block} sizes were configured for kernel execution. Finally, each particular test was run twenty times and the performance was calculated as their average to avoid variability.

\begin{table}[tb]
\centering
\caption{Experimental platforms used in the tests}
\label{tab:platforms}
\resizebox{\textwidth}{!}{%
\begin{tabular}{@{}lllllll@{}}
\toprule
\multicolumn{3}{c}{\textbf{CPU}}        & \multicolumn{4}{c}{\textbf{GPU}}            \\
\multicolumn{1}{c}{\textbf{ID}} & \multicolumn{1}{c}{\textbf{Processor}} & \multicolumn{1}{c}{\textbf{\begin{tabular}[c]{@{}c@{}}RAM \\ Memory\end{tabular}}} & \multicolumn{1}{c}{\textbf{ID}} & \multicolumn{1}{c}{\textbf{\begin{tabular}[c]{@{}c@{}}Vendor\\ (Type)\end{tabular}}} & \multicolumn{1}{c}{\textbf{\begin{tabular}[c]{@{}c@{}}Model\\ (Architecture)\end{tabular}}} & \multicolumn{1}{c}{\textbf{\begin{tabular}[c]{@{}c@{}}GFLOPS\\ peak (SP)\end{tabular}}} \\ \midrule

\textit{Core-i5} & \begin{tabular}[c]{@{}l@{}}Intel Core \\ i5-7400\end{tabular} & 16 GB & \textit{Titan}
& \begin{tabular}[c]{@{}l@{}}NVIDIA\\ (Discrete)\end{tabular} & \begin{tabular}[c]{@{}l@{}}Titan X\\ (Pascal)\end{tabular} & 10970 \\
\textit{Core-i3} & \begin{tabular}[c]{@{}l@{}}Intel Core\\ i3-4160\end{tabular} & 8 GB & \textit{RTX} & \begin{tabular}[c]{@{}l@{}}NVIDIA\\ (Discrete)\end{tabular} & \begin{tabular}[c]{@{}l@{}}RTX 2070\\ (Turing)\end{tabular} & 7465 \\

\textit{Core-i9} &  \begin{tabular}[c]{@{}l@{}}Intel Core \\ i9-10920X\end{tabular} & 32 GB & \textit{Iris XE} & \begin{tabular}[c]{@{}l@{}}Intel \\ (Discrete)\end{tabular} & \begin{tabular}[c]{@{}l@{}}Iris Xe MAX Graphics\\ (Gen 12.1)\end{tabular} & 2534 \\
\textit{Xeon} & \begin{tabular}[c]{@{}l@{}}Intel Xeon \\ E-2176G\end{tabular} & 65 GB  & \textit{P630} & \begin{tabular}[c]{@{}l@{}}Intel\\ (Integrated)\end{tabular} & \begin{tabular}[c]{@{}l@{}}UHD Graphics P630\\ (Gen 9.5)\end{tabular} & 441.6 \\

 \bottomrule
\end{tabular}
}
\end{table}

\subsection{Performance Results}

\begin{figure}[t!]
 \centering
 \begin{minipage}[b]{.48\textwidth}
     \centering
      \includegraphics[width=1\columnwidth,scale=1,bb=0 0 600 371]{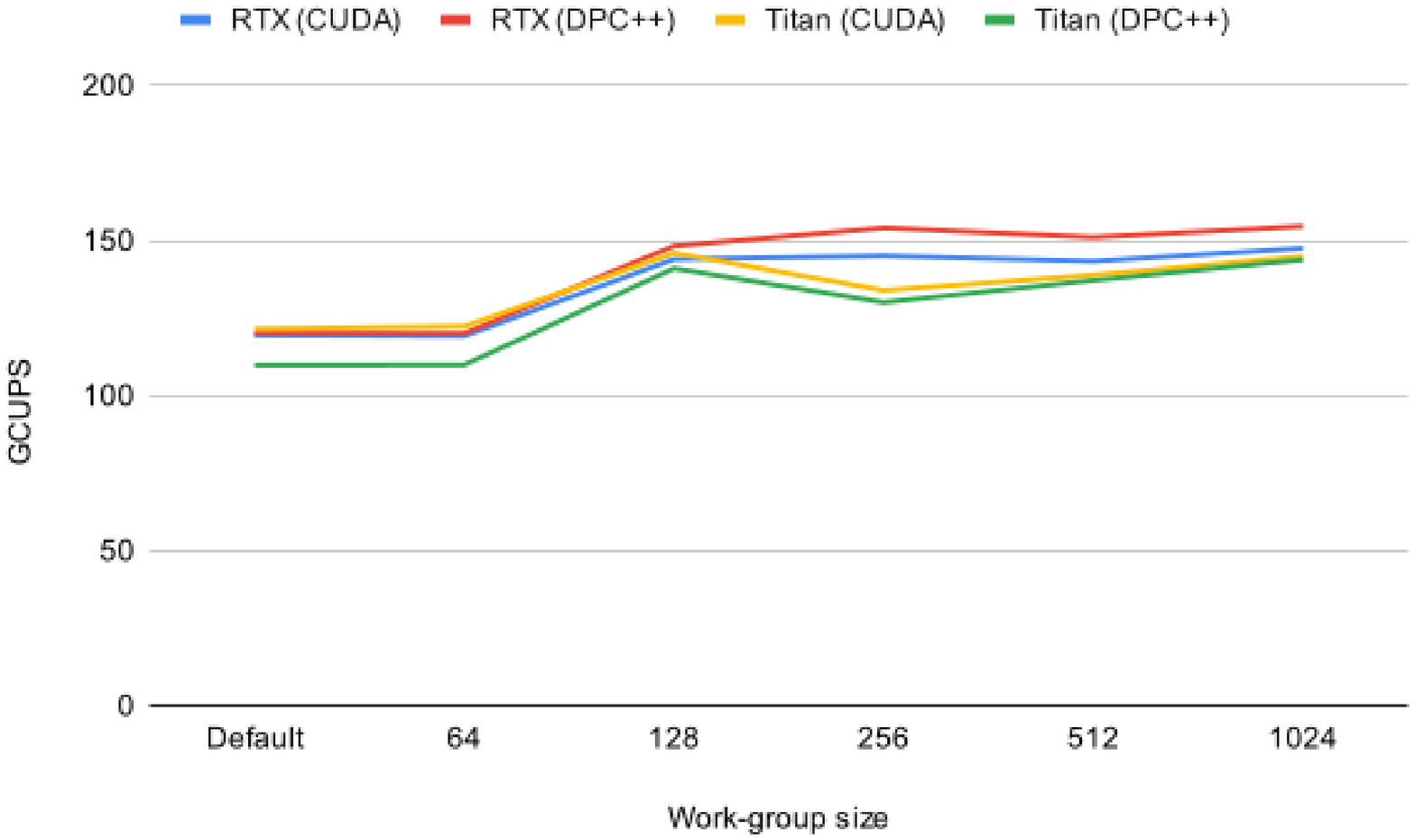}
     \caption{Performance of both CUDA and DPC++ versions on the NVIDIA GPUs when varying work-group size.}
     \label{nvidia-gpus-x20}
 \end{minipage}\hfill%
 \begin{minipage}[b]{.48\textwidth}
     \centering
      \includegraphics[width=1\columnwidth,scale=1,bb=0 0 600 371]{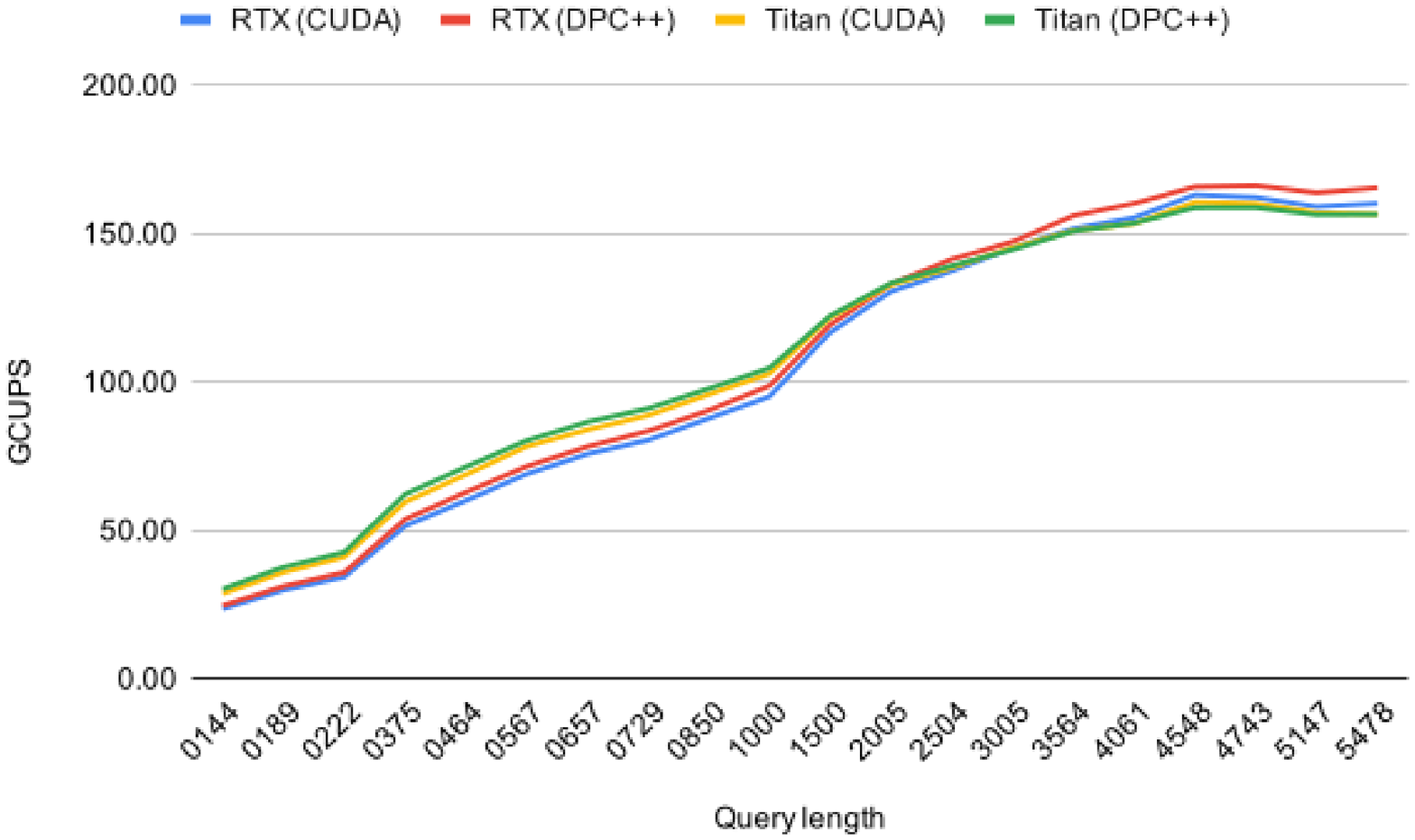}
     \caption{Performance of both CUDA and DPC++ versions on the NVDIA GPUs when varying the query lenght.}
     \label{nvidia-gpus-query}
 \end{minipage}
 \end{figure}

GCUPS (billion cell updates per second) is commonly used as the performance metric in the context of SW~\cite{OSWALD16}. Fig.~\ref{nvidia-gpus-x20} presents the performance of both CUDA and DPC++ versions on two  NVIDIA  GPUs  when  varying work-group size. First, it can be noted that both codes are sensitive to the work-group size. In fact, the best performances are reached using work-group sizes different to the ones that SW\#db set as default. Regarding the performance on each NVIDIA GPU, there are no significant differences between both codes on the Titan. However, on the RTX, this situation gets reversed; the DPC++ version outperforms its CUDA counterpart (approximately 5\%).


Fig.~\ref{nvidia-gpus-query} deepens the above analysis presenting the performance of both CUDA and DPC++ versions on the  NVIDIA  GPUs  when  varying the query length (optimal work-group size is used for each case). It can be noted that all versions benefit from larger workloads. As expected, the CUDA code achieves the same GCUPS as the DPC++ one for all query lengths on the Titan. Both DPC++ and CUDA versions present practically the same performance on the RTX, outperforming the latter to the former on the largest sequences.



\begin{figure}[t!]
 \centering
 \begin{minipage}[b]{.48\textwidth}
     \centering
      \includegraphics[width=1\columnwidth,scale=1,bb=0 0 600 371]{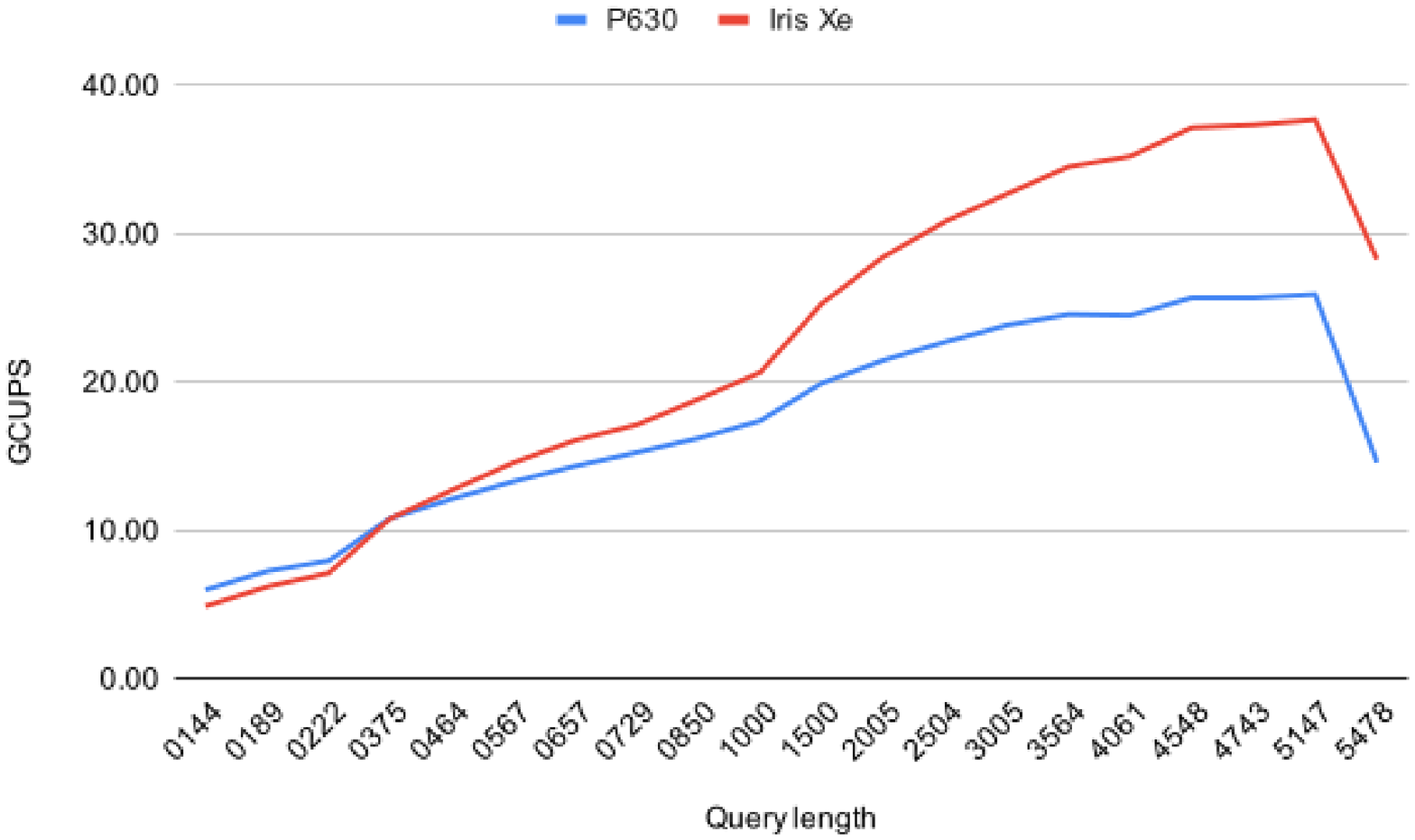}
     \caption{Performance of DPC++ code on the different Intel GPUs when varying the query lenght.}
     \label{intel-gpus}
 \end{minipage}\hfill%
 \begin{minipage}[b]{.48\textwidth}
     \centering
      \includegraphics[width=1\columnwidth,scale=1,bb=0 0 600 371]{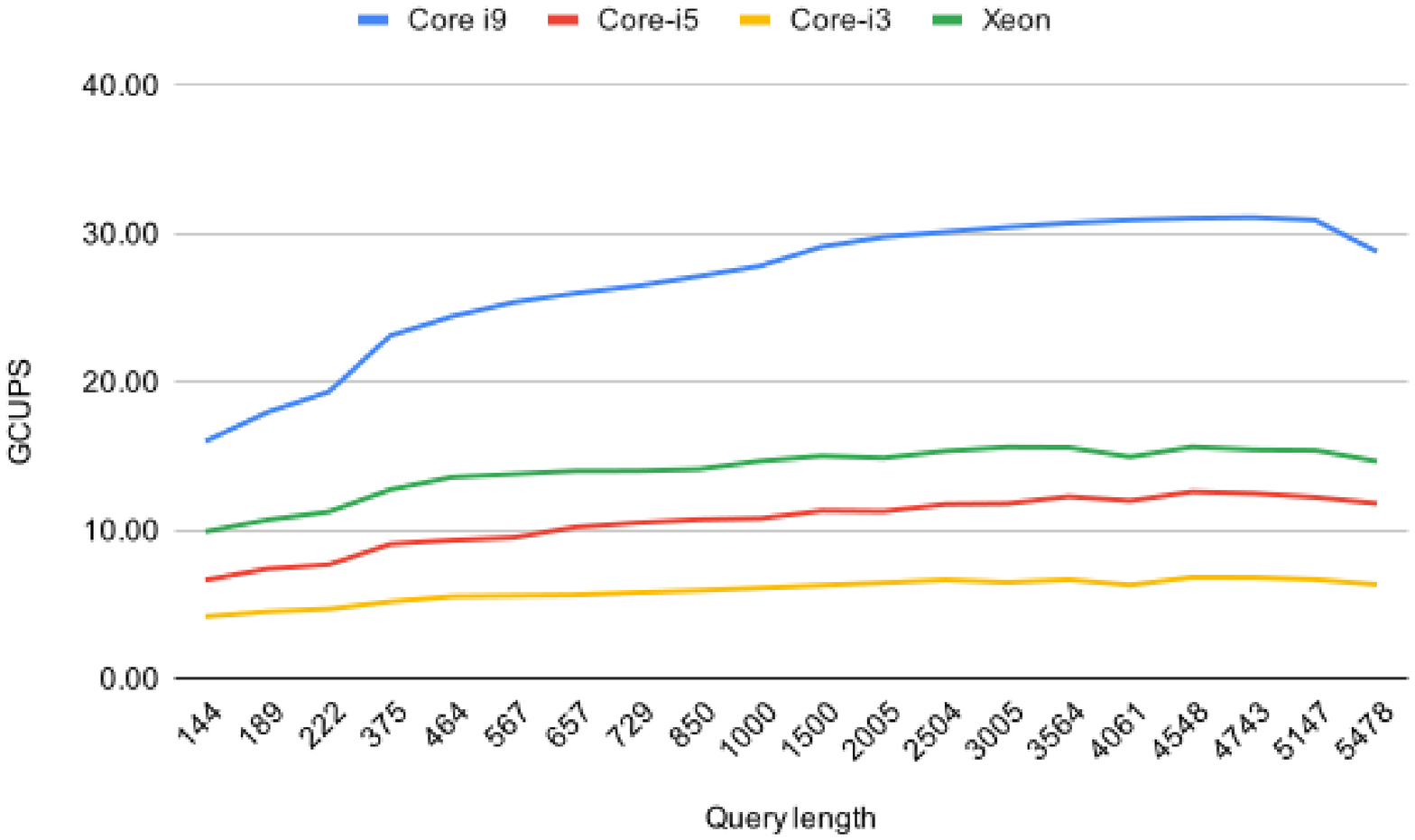}
     \caption{Performance of DPC++ code on the different Intel CPUs when varying the query lenght.}
     \label{intel-cpus}
 \end{minipage}
 \end{figure}

To verify cross-GPU vendor portability, the DPC++ code was executed on two different Intel GPUs varying the query lenght (see Fig.~\ref{intel-gpus}). Due to the
absence of an optimized version for both Intel devices, little can be said about its performance. However, it is important to remark that only two minor changes were necessary to carry out these tests: (1) setting the appropriate work-group size and (2) setting the corresponding backend. As the ported code was compiled and executed with minimal tuning, there is probably room for further improvement.

Finally, Fig.~\ref{intel-cpus} presents the performance of the DPC++ code on 4 different Intel CPUs, demonstrating its cross-architecture portability. Considering performance, more GCUPS are reached as the query length increases. Once again, running the migrated code just required minimal intervention and its performance could be improved through fine tuning.

\section{Conclusions and Future Work}
\label{sec:conc}

The recently introduced Intel oneAPI ecosystem aims to respond to the programming challenge related to heterogeneous computing.
In this paper, we present our experiences migrating a CUDA-based, biological software tool to DPC++ using the oneAPI framework. 
Among the main contributions of this research we can summarize:

\begin{itemize}
    \item \texttt{dpct} proved to be an effective tool for SW\#db code migration to DPC++. While it was not able to translate the complete code, \texttt{dpct} did most of the work and gave hints to the programmer on the pending parts.
    \item The migrated code could be successfully executed on CPUs and also GPUs from different vendors, demonstrating its cross-architecture, cross-GPU vendor portability.
    \item Performance results showed that the migrated DPC++ code is comparable to the original CUDA one. In fact, DPC++ can be even faster in some cases. As the ported code was compiled and executed with minimal tuning, there is probably room for further improvement.


\end{itemize}

Future work will focus on:

\begin{itemize}
    \item Understanding the gap in performance between DPC++ and CUDA codes, and optimizing DPC++ codes to reach their maximum performance.
    \item Carrying out more exhaustive experimental work. In particular, considering other alignment operations, larger workloads, multi-GPU execution, among others, to increase the representativeness of this study.
    \item Running the DPC++ code on other architectures like FPGAs, to verify its cross-architecture portability.
\end{itemize}

\section*{Acknowledgements}

This paper has been supported by the EU (FEDER) and the Spanish MINECO and CM under grants S2018/TCS-4423, RTI2018-093684-B-I00 and PID2021-126576NB-I00.


%
%
%
\bibliographystyle{splncs04}
\bibliography{references}
\end{document}